\documentclass[aps, prl,floats, floatfix, twocolumn,superscriptaddress, footinbib, showpacs]{revtex4-1}

\usepackage{graphicx}
\usepackage{hyperref}
\usepackage{amsmath,amssymb}
\usepackage{amsfonts}
\usepackage{xspace} 
\usepackage[usenames]{color}
\usepackage{dcolumn} 
\usepackage{bm} 
\usepackage{epsfig}
\usepackage{aas_macros}

\usepackage{ulem}
\definecolor {darkgreen}{rgb}{0.2,0.7,0.2}

\newcommand{\be}{\begin{equation}}
\newcommand{\ba}{\begin{eqnarray}}
\newcommand{\ee}{\end{equation}}
\newcommand{\ea}{\end{eqnarray}}

\newcommand{\SMBH}{\bullet}
\newcommand{\SCO}{\star}
\newcommand{\GW}{{\mbox{\tiny GW}}}
\newcommand{\vac}{{\mbox{\tiny vac}}}

\newcommand{\Msun}{\,{\rm M_\odot}}
\newcommand{\yr}{\,{\rm yr}}

\newcommand{\D}{\mathrm{d}}

\newcommand{\R}{{\bar{r}}}

\begin{document}

\title{Imprint of Accretion Disk-Induced Migration on Gravitational Waves from Extreme Mass Ratio Inspirals}

\author{Nicol\'as Yunes}
\affiliation{Dept.~of Physics and MIT Kavli Institute, 77 Massachusetts Avenue, Cambridge, MA 02139.}
\affiliation{Harvard-Smithsonian Center for Astrophysics, 60 Garden St., Cambridge, MA 02138, USA.}

\author{Bence Kocsis}
\affiliation{Harvard-Smithsonian Center for Astrophysics, 60 Garden St., Cambridge, MA 02138, USA.}

\author{Abraham Loeb}
\affiliation{Harvard-Smithsonian Center for Astrophysics, 60 Garden St., Cambridge, MA 02138, USA.}

\author{Zolt\'an Haiman}
\affiliation{Department of Astronomy, Columbia University, 550 West 120th Street, New York, NY 10027}

\begin{abstract}
We study the effects of a thin gaseous accretion disk on the inspiral of a stellar--mass black hole into a supermassive black hole. We construct a phenomenological angular momentum transport equation that reproduces known disk effects. Disk torques modify the gravitational wave phase evolution to detectable levels with LISA for reasonable disk parameters. The Fourier transform of disk-modified waveforms acquires a correction with a different frequency trend than post-Newtonian vacuum terms. Such inspirals could be used to detect accretion disks with LISA and to probe their physical parameters.
\end{abstract}
\date{\today \hspace{0.2truecm}}
\maketitle

The inspiral of a stellar-mass compact object (SCO), such
as a black hole or neutron star, into a supermassive black
hole (SMBH) is among the most interesting gravitational wave (GW)
sources for the Laser Interferometer Space Antenna (LISA)~\cite{lisa}.
Extreme mass ratio inspirals (EMRIs) can be produced
through fragmentation of accretion disks
into massive stars~\cite{2004ApJ...608..108G,Levin:2006uc}
or through capture of compact remnants
by hydrodynamic drag \cite{1999A&A...352..452S,*2005ApJ...619...30M},
which are believed to be mass-segregated in galactic
nuclei~\cite{2006ApJ...649...91F},
as well as through other channels~\cite{2007CQGra..24..113A}. Stars
which reside within an accretion disk will lead to EMRIs, provided
they become a SCO in less time than their inward migration time.
Although the expected EMRI event rate is rather uncertain
(between a few tens to hundreds over LISA's lifetime, including
coalescences and inspiral-only events~\cite{Gair:2008bx}), a
detectable fraction may originate in active galactic nuclei (AGNs)
with an accretion disk.

Accretion disks are efficient at extracting orbital angular momentum
from the extreme mass-ratio binary. The SCO torques the disk gravitationally,
inducing spiral density waves that carry away angular
momentum~\cite{2007astro.ph..1485A,*1997Icar..126..261W}. In planetary disks, the same phenomenon
leads to {\emph{migration}} of planets towards their parent star.
Planetary migration has been classified into different types (determined by disk parameters,
the EMRI's mass ratio and orbital separation) to distinguish
circumstances where a gap opens around the planet (Type-II) from those without a
gap (Type-I). In EMRIs, migration becomes the dominant source of angular
momentum transport at separations $\gtrsim 100 M_{\SMBH}$, where
$M_{\SMBH}$ is the SMBH mass~\cite{2001A&A...376..686K,Levin:2006uc}
(we use units $G = c = 1$).

Migration changes the relation between the binary's binding energy and
the GW luminosity, and hence it affects the inspiral rate and the GW phase
evolution. EMRIs enter the LISA sensitivity band only inside $\lesssim
50 M_{\SMBH}$, where GW angular momentum transport is dominant. Thus,
migration acts perturbatively in LISA EMRIs. In this Letter, we
examine whether the imprint of migration on the EMRI GW observables is
detectable by LISA. In a companion paper~\cite{long-paper},
we consider a broader range of disk effects and their impact on GWs in more detail.

{\emph{Disk Properties and Migration.}---We consider
radiatively-efficient, geometrically thin accretion disks, whose two most
important free parameters are the accretion rate $\dot{M}_{\SMBH}$
(overhead dots denote time-derivatives) and the
$\alpha$-viscosity parameter. AGN observations suggest an accretion
rate $\dot{M}_{\SMBH} \equiv \dot m_{\SMBH} \dot{M}_{\SMBH\rm Edd} \in
(0.1,1) \dot{M}_{\SMBH\rm Edd}$~\cite{2006ApJ...648..128K}. Evidence
for the magnitude of $\alpha$ is inconclusive, with plausible
theoretical and observed ranges in
$(0.01,1)$~\cite{2007ApJ...668L..51P,*2001A&A...373..251D,*2007MNRAS.376.1740K}.
We focus on Shakura-Sunyaev $\alpha$-disks~\cite{1973A&A....24..337S} and
$\beta$-disks~\cite{1981ApJ...247...19S}, which differ in whether viscosity
is proportional to the total pressure (gas plus radiation) or only the gas
pressure, respectively. This affects the surface density
($\Sigma \propto r^{3/2}$ and $r^{-3/5}$ for $\alpha$- and
$\beta$-disks when opacity is dominated by  electron scattering).
The local disk mass is much larger for $\beta$--disks at radii $r\ll 10^{3} M_{\SMBH}$,
leading to a stronger GW imprint.

In the absence of a gap, Type-I migration models for angular momentum
transport have been formulated~\cite{1980ApJ...241..425G,2002ApJ...565.1257T}  but they
are very sensitive to opacity and
radiation processes~\cite{2006A&A...459L..17P} and lack the
stochastic features observed in magnetohydrodynamic
simulations~\cite{2004MNRAS.350..849N,*2004ApJ...608..489L}.  The
presence of a gap leads to Type-II models for
angular momentum exchange~\cite{1995MNRAS.277..758S,1999MNRAS.307...79I}.
These also oversimplify the process, assuming either a steady state or
quasi-stationarity. Type-II migration can also cease interior to a decoupling radius, $r_{\rm d}$,
in the late stages of the inspiral, when the gas accretion velocity outside the gap
becomes slower than the SCO's GW-driven inspiral velocity~\cite{2005ApJ...622L..93M,*2009ApJ...700.1952H}.
Alternatively, the gap can refill by non-axisymmetric or 3D inflow, restoring viscous torque
balance from inside and outside the SCO's orbit and slowing the gaseous
migration~\cite{1996ApJ...467L..77A,*2008ApJ...672...83M,*2009MNRAS.393.1423C}.
Migration is mostly unexplored in the regime relevant to LISA EMRIs,
i.e.~for radiation-pressure dominated, optically-thick, geometrically-thin, relativistic,
magnetized and turbulent disks, with the SCO's mass $m_{\SCO}$
exceeding the local disk mass.

Astrophysical uncertainties regarding accretion disks and migration in
the regime relevant to LISA EMRIs lead us to consider a general
power-law relation,
\ba
\label{eq:L}
\dot{\ell}_{\SCO} &=& \dot{\ell}_{\GW} \left(1 + \delta \dot{\ell}\right)\,,
\\
\delta \dot{\ell} &\equiv& A \; \R^{B} =
A_{0} \;
\alpha_{1}^{A_{1}} \;
\dot{m}_{\SMBH 1}^{A_{2}} \;
M_{\SMBH 5}^{A_{3}} \;
m_{\SCO 1}^{A_{4}} \;
\R^{B}\,,
\label{eq:delta}
\ea
where $\dot{\ell}_{\SCO}$ is the SCO's rate of change of specific
angular momentum, $\dot{\ell}_{\GW}$ is the loss due to GWs,
and $\delta \dot{\ell}$ a correction induced by
migration. The power-law form in the reduced radius $\R \equiv
r/M_{\SMBH}$ involves an amplitude $A$, which is parameterized in
terms of normalized accretion disk ($\alpha_{1} \equiv
\alpha/0.1$, $\dot m_{\SMBH} = \dot m_{\SMBH}/0.1$),
and mass parameters $[M_{\SMBH 5} = M_{\SMBH}/(10^{5} \Msun)$, $m_{\SCO 1} = m_{\SCO}/(10 \Msun)$].
The power-law indices $(A_{i>0},B)$ are given in Table~\ref{flux-Table-pars}
for representative migration models: rows 1--2 correspond to
Type I \cite{2002ApJ...565.1257T}, 2--4 to steady state Type-II \cite{1995MNRAS.277..758S},
5 to quasistationary Type-II migration in the asymptotic limit for small
$\R$ \cite{1999MNRAS.307...79I} (the latter is available for $\beta$--disks only).
The gap decouples and Type-II migration ceases ($A\approx 0$) interior to
$\R_{\rm d} = 1.4 \times 10^{-5} \alpha_{1}^{-2} \dot{m}_{\SMBH 1}^{-4} M_{\SMBH
 5}^{-2} m_{\SCO 1}^{2} \lambda^{5}$ for $\alpha$ and 
$15\,\alpha_{1}^{-4/13} \dot{m}_{\SMBH 1}^{-2/13}
 M_{\SMBH 5}^{-4/13}  m_{\SCO 1}^{5/13} \lambda^{2/13}$
for $\beta$ disks, where $\lambda r$ is the gap radius (we adopt
 $\lambda = 1.7$ \cite{1994ApJ...421..651A}).
Since disk effects become stronger at larger radii, $B>0$.
\begin{table}
\begin{ruledtabular}
\begin{tabular}{ccccccccc}
 & & $A_{0}$ & $A_{1}$ & $A_{2}$ & $A_{3}$ & $A_{4}$ & $B$  \\
\hline
\cite{2002ApJ...565.1257T}, I$\alpha$ & {\color{Red}\Large$\bullet$} 	& $7.2 \times 10^{-19}$ & $-1$   &  $-3$ & $1$ & $0$ &  $8$\\
\cite{2002ApJ...565.1257T}, I$\beta$ & {\color{Blue}\Large${\bullet}$} 	& $6.5 \times 10^{-13}$ & $-4/5$ & $-7/5$ & $6/5$ & $0$ & $59/10$ \\
\cite{1995MNRAS.277..758S}, II$\alpha$ & ${\color{Red}\Box}$ & $6.2 \times 10^{-10}$ & $0$  & $1$ & $3$ & $-2$ & $4$ \\
\cite{1995MNRAS.277..758S}, II$\beta$ & ${\color{Blue}\Box}$ 	& $4.4 \times 10^{-6}$ & $1/2$ & $5/8$ &  $13/8$ & $-11/8$ & $25/8$\\
\cite{1999MNRAS.307...79I}, II$\beta$ & ${\color{Blue}\bigtriangleup}$ 	& $1.6 \times 10^{-7}$ & $2/7$ & $11/14$ &  $31/14$ & $-23/14$ & $7/2$
\end{tabular}
\caption{\label{flux-Table-pars}
Disk parameters for Type I and II migration models in $\alpha$ and $\beta$ disks.}
\end{ruledtabular}
\end{table}

{\emph{GW Implications.}---}The change in the angular momentum dissipation rate due to migration modifies the GW evolution, leading to a change in the accumulated GW phase and spectrum.
For circular orbits, the quadrupolar GW phase can be computed from $\phi_{\GW} = 2\int_{r_{0}'}^{r_{f}} dr\, {\Omega} \, {\dot{\ell}_{\SCO}}^{-1} \, {d\ell}/{dr}$, where the orbital frequency is $\Omega \simeq (M_{\SMBH}/r^{3})^{1/2}$, the binary's specific angular momentum is $\ell=r^2 \Omega = M_{\SMBH}^{1/2} r^{1/2}$, while the specific angular momentum flux $\dot{\ell}_{\SCO}$ is given by Eq.~\eqref{eq:L}. For a fixed final EMRI separation $r_f$ and observation time $T_{\rm obs}$, the initial separation $r_{0}'$ is different from what it would be in vacuum, as the radial inspiral evolution $\dot{r}$ is determined by $\dot{\ell}_{\SCO}$: $r = \int_{r_{0}'}^{r_{f}} \dot{\ell}_{\SCO} (d\ell/dr)^{-1} dr$. For an unperturbed EMRI, $(\R_{f}/\R_{0})^{-4} \approx 1 + 33\, (m_{\SCO 1}/M_{\SMBH 5}^{2}) (T_{\rm obs}/{\rm yr}) (\R_{f}/10)^{-4}$.

The correction to the GW phase, $\delta \phi_{\GW} \equiv \phi_{\GW} - \phi_{\GW}^{\vac}$, where $\phi_{\GW}^{\vac}$ is the accumulated phase in vacuum, is then
\begin{eqnarray}
\label{gen-form}
\delta \phi_{\GW} &=&
\bar{A} \;  \frac{M_{\SMBH 5}}{m_{\SCO 1}} \;
\R_{0}^{B+5/2}
\\ \nonumber
&&
\times\left(1 + \frac{2B+5}{3} x^{B+4}
 -\frac{2B+8}{3}  x^{B+5/2}  \right)\,,
\end{eqnarray}
where $\bar{A} \equiv -(3 \times 4^{-1/2} \times 5^{5}) (4 + B)^{-1} (5 + 2 B)^{-1} A$,
$x \equiv r_{f}/r_{0}$, $\R_{0} \equiv r_{0}/M_{\SMBH}$ and
we have expanded in $\delta \dot{\ell} \ll 1$, which holds in the LISA regime.
For fixed $T_{\rm obs}$, we find that $|\delta \phi_{\GW}|$  increases and decreases with $m_{\SCO}$ for
the Type-I and II models of Table~\ref{flux-Table-pars}, respectively.

\begin{figure}[th]
\centering
 \includegraphics[width=8.5cm,clip=true]{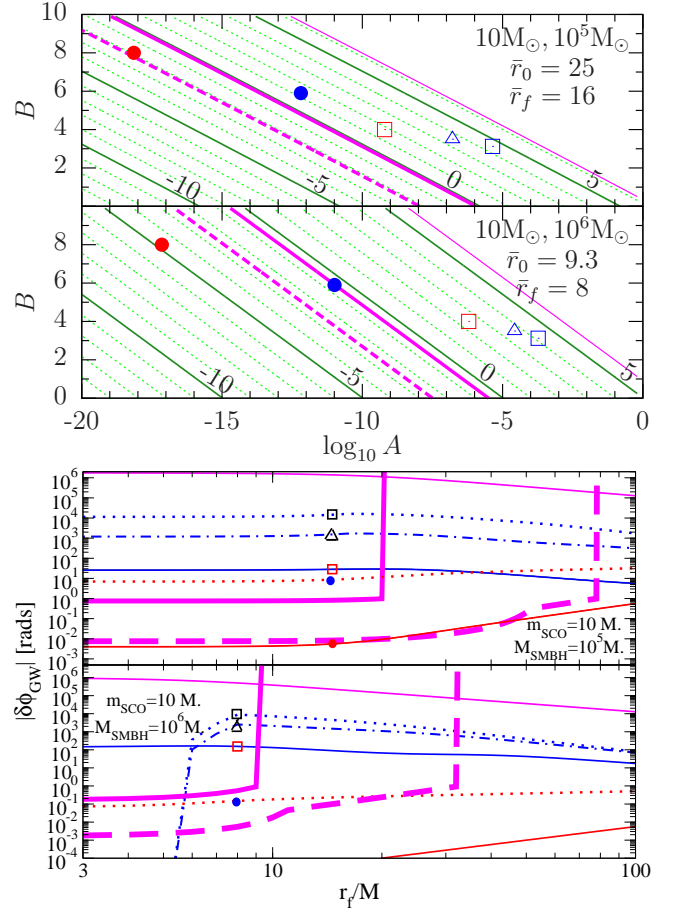}
\\\includegraphics[width=8.5cm,clip=true]{analytic-dephasing-prl.eps}
 \caption{\label{fig:analytic-dephasing} Top: contours of $\log_{10}|\delta \phi_{\GW}|$
 for different interaction models of two EMRI systems observed for one year.
Bottom: $|\delta \phi_{\GW}|$ for different $\R_{f}$ for the 5 models in Table~\ref{flux-Table-pars}.
 Many of the models -- especially those resembling $\beta$-disks and Type II migration --
 shown in Table~\ref{flux-Table-pars} (marked with symbols in the top, and
 individual curves on the bottom) lie well above the LISA sensitivity level (thick magenta lines).
}
\end{figure}
The top panel of Fig.~\ref{fig:analytic-dephasing} plots the dephasing in
Eq.~\eqref{gen-form} for two typical LISA EMRIs at fixed $r_f>r_{\rm d}$ as contours for different torque parameters $(A,B)$.
The specific migration models defined in Table~\ref{flux-Table-pars} with $\alpha_{1} = 1 = \dot{m}_{\SMBH 1}$
are marked with symbols. The bottom panel shows $\delta \phi_{\GW}$ for those models but with different $r_f$,
fixing $T_{\rm obs}=1\yr$ (c.f. LISA's planned lifetime is 3 years).
For comparison, we also plot the total GW phase accumulation [${\cal O}(10^6)$ top, thin line]
and a rough measure of LISA's accuracy to phase measurements: $\delta \phi_{\GW} >
10/\rho$, where $\rho$ is the signal to noise ratio $\rho(h)=4\int_{r_{0}}^{r_{f}} \D
r (df/dr) |\tilde{h}|^2 S_n^{-1}[f(r)]$, with $S_n[f(r)]$ the
LISA detector noise~\cite{2004PhRvD..70l2002B} and $\tilde{h}$ the Fourier transform
of the orientation-averaged GW signal. We evaluate $\rho$ at 1 Gpc (or redshift $z\approx
0.2$; thick solid line) and at 10 Mpc (or $z \approx 0.002$;
thick dashed line). For $\rho<10$, we assume the EMRI is not detected at all, which explains the sharp rise in the
detection level beyond a certain $r_{f}$. Migration with a gap (empty symbols) causes a bigger
phase shift because of the pileup of mass outside the gap.
For $\R_{\rm d} \lesssim \R_f \lesssim 50$ but fixed $(A,B,M_{\SMBH},m_{\SCO},T_{\rm obs})$,
the phase shift is constant within a factor $\sim 3$,
but it quickly drops off for the Type-II models interior to the gap
decoupling radius $r_{\rm d}$ where $A\rightarrow 0$.

The Newtonian estimates presented here suggest that LISA EMRI
observations might be able to probe accretion-disk induced
migration. Figure~\ref{fig:analytic-dephasing} shows that a large
sector of parameter space $(A,B)$
exists where the dephasing is large enough to be detectable, and
$\delta\phi_{\GW}$ is very sensitive to the disk model and its parameters.
One might worry, however, that the estimates in
Fig.~\ref{fig:analytic-dephasing} are inaccurate due to the use of a
Newtonian waveform model. We have verified that this is not the case
through a relativistic waveform model that employs the calibrated
effective-one-body scheme~\cite{Yunes:2009ef,*2009GWN.....2....3Y,*Yunes:2010zj}.
We have generated 1 year-long waveforms for the systems plotted in Fig.~\ref{fig:analytic-dephasing}
and included modifications to the radiation-reaction force due to
migration, as parameterized by Eq.~\eqref{eq:delta}. Overall, we
find the Newtonian results to be representative of the fully
relativistic ones~\cite{long-paper}.

Just because migration produces a sufficiently large phase correction
does not necessarily imply that LISA can measure it. For that to be
possible, migration phase corrections must be non-degenerate, or at
worst, weakly correlated with other system parameters. One can
study if this is the case by computing the Fourier transform of
the GW observable. We employ the stationary phase approximation
(SPA)~\cite{1995PhRvD..52..848P,*Yunes:2009yz}, where one assumes the GW phase
varies much more rapidly than the amplitude. The Fourier transform of
$h(t) = A(t) \exp[i \phi_{\GW}(t)]$ can then be approximated as
$\tilde{h}(f) = \tilde{A}(f) \exp[i \psi(f)]$, where $\tilde{A}(f)
\equiv (4/5) A[t(f)] {\dot{f}}^{-1/2}$ and $\psi(f) \equiv 2 \pi f
t_{0} - \phi_{\GW}(t_{0})$, where $f$ is the GW frequency and $t_{0}$
is the stationary point, defined by
$2 \pi f = (d\phi_{\GW}/dt)_{t=t_{0}}$~\cite{1995PhRvD..52..848P,*Yunes:2009yz}.

The corrections due to migration on the Fourier transform of the GW phase in the SPA,
$\delta \psi \equiv \psi - \psi_{\rm vac}$, are
\begin{align}
\frac{\delta \psi}{\psi_{\rm vac}^{\rm Newt}} =
\tilde{A} \;
\bar{\eta}^{2B/5} \;
\bar{u}^{-2 B/3}\,,
\label{FT-SPA-correction}
\end{align}
where we have defined $\tilde{A} \equiv -2^{2-8B/5}
5^{1-8B/5} (4+B)^{-1} (5+2B)^{-1} A \exp(6.46 B)$, the normalized symmetric mass ratio
$\bar{\eta} \equiv m_{\SCO1}/{M_{\SMBH5}}$ and $\bar{u} \equiv (\pi
{\cal{M}} f)/(6.15 \times 10^{-5})$, and where ${\cal{M}} =
m_{\SCO}^{3/5} M_{\SMBH}^{2/5}$ is the chirp mass and $\psi_{\rm
vac}^{\rm Newt} = (3/128) (\pi {\cal{M}} f)^{-5/3}$ is the
leading-order (Newtonian) vacuum Fourier phase. The amplitude of the
SPA Fourier transform is corrected in a similar fashion: $\delta
|\tilde{h}|/|\tilde{h}|_{\rm vac} \sim \delta \psi/\psi_{\rm
vac}$.

Equation~\eqref{FT-SPA-correction} is to be compared with the
intrinsic general relativity (GR) corrections to the vacuum Fourier GW phase:
$\psi_{\rm vac}/{\psi_{\rm vac}^{\rm Newt}} =
\sum_{n=0}^{\infty} a_{n} u^{2n/3}$, where
$a_{q}=a_{q}(m_{\SCO},M_{\SMBH})$,
and the modulation induced by the orbital motion of LISA around the Sun.
Migration corrections lead to
negative frequency exponents in the Fourier phase (in
Eq.~\eqref{FT-SPA-correction}, $-2B/3 < 0$), while GR, post-Newtonian
corrections in vacuum lead to positive powers of
frequency, while the detector orbit is periodic with a $1\yr$ period.
For a sufficiently strong signal, this suggests it might be possible to separate the migration
effects from the other GR and detector orbit induced phase corrections.

Modified gravity theories might introduce corrections
similar to Eq.~\eqref{FT-SPA-correction}. The parameterized
post-Einsteinian (ppE) framework~\cite{Yunes:2009ke}, devised to search for
generic GR deviations in GW data, postulated such a phase modification,
allowing for both positive and negative $B$. Degeneracies between disk and
modified gravity effects with negative frequency exponents ($B>0$)
could then exist (e.g.~Brans-Dicke theory or $G(t)$ theories).
The latter, however, have already been greatly constrained by binary
pulsar observations~\cite{Yunes:2009bv}. Moreover, alternative theory
modifications should be present in all EMRIs, while disk effects will be present
in only a small subset.

A precise measure of whether a migration-modified waveform
$\tilde{h}_1(f)$ is distinguishable from a vacuum waveform
$\tilde{h}_{2}(f;\vec{\lambda})$, where $\vec{\lambda}$ stands for all
disk parameters, requires a detailed Monte Carlo study
that maps the likelihood surface. A rough measure of
distinguishability can be obtained by calculating the signal-to-noise ratio (SNR)
of the difference between a vacuum and a non-vacuum waveform $\rho(\delta h)$
by minimizing only over a time and a phase shift. Using this crude measure,
we demonstrate in the bottom panel of Fig.~\ref{fig:SNR-diff} that most of the migration
models of Table~\ref{flux-Table-pars} lead to $\rho(\delta h)>10$ within $4$ months of
observation for a source at $1$ Gpc.

\begin{figure}[th]
 \includegraphics[width=8.5cm,clip=true]{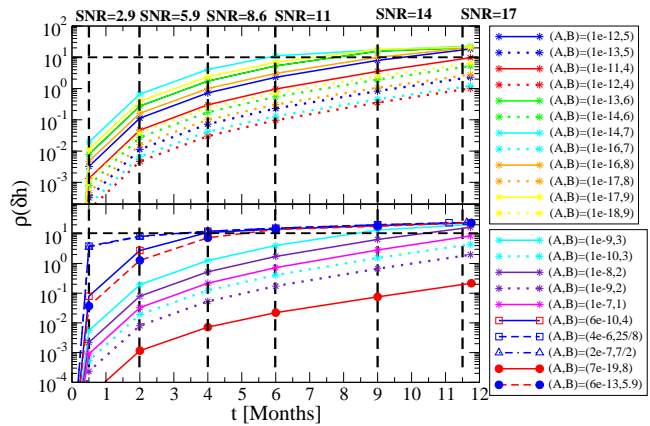}
 \caption{\label{fig:SNR-diff} $\rho(\delta h)$ as a function of observation time.
Observe that $\rho(\delta h)>5$ within 1 year for a large set of $(A,B)$.}
\end{figure}
Going beyond Table~\ref{flux-Table-pars}, there exists a
large sector of disk parameter space $(A,B)$ for which the SNR of the
waveform difference exceeds threshold $\rho(\delta h) > 10$.
Figure~\ref{fig:SNR-diff} plots $\rho(\delta h)$ at $1$ Gpc as a function
of observation time for different values of $(A,B)$ in Eq.~\eqref{eq:delta}.
We also indicate (with labels over the vertical dashed lines)
the SNR of the vacuum waveform at 1 Gpc and
$T=(0.5,2,4,6,9,11.5)$ months. We calculate the waveforms with the relativistic
model of~\cite{Yunes:2009ef,*2009GWN.....2....3Y,*Yunes:2010zj}
and $(M_{\SMBH 5},m_{\SCO 1}) = (1,1)$, SMBH spin angular momentum
$|S_{\SMBH}| = 0.9 M_{\SMBH}^{2}$ coaligned with the orbital angular momentum,
and initial and final separations $(r_{0},r_{f}) \sim (24.5,16) M_{\SMBH}$.
Observe that for a large set of disk parameters $(A,B)$, $\rho(\delta h) > 10$
within a one-year observation. Fitting to the smallest $A$ with $\rho(\delta h)>10$ for fixed $B$,
we find that for these masses and orbital radii, LISA could measure
$\log_{10} A \gtrsim a_{1} + a_{2} B$, with $a_{1} = -5.7 \pm 0.4$
and $a_{2}=-1.4 \pm 0.2$.

{\emph{Discussion.}---The GW observation of EMRI signals with LISA
could be used to probe the uncertain physics of accretion disks. In
particular, spiral density waves generated by an orbiting SCO can
transfer sufficient orbital angular momentum to alter the GW signal at
levels that are detectable by LISA. The effects are strongest for parameter choices 
resembling Type II migration (i.e. with the SCO opening a gap) in relatively massive 
$\beta$ disks. A very crude (diagonal) Fisher analysis
suggests that LISA could measure certain sectors of disk parameter space
to better than $10\%$, for vacuum SNRs larger than $10$, the details of
which will be presented in a companion paper~\cite{long-paper}.
This is no surprise considering that  $\delta \phi_{\GW}$ is at worst $\sim 10$
times higher than LISA's sensitive curve in Fig.~\ref{fig:analytic-dephasing}.
Detection of the predicted migration effect would reduce the uncertainty in existing
theoretical models and offer the potential for extending the
discussion to more complicated geometries (such as EMRIs with
eccentric and/or inclined orbits).

The detection of EMRIs in AGNs and the extraction of
disk-parameters improve the prospects for finding electromagnetic
counterparts in the LISA error volume with consistent
luminosities~\cite{2006ApJ...637...27K,*2008ApJ...684..870K}. Coincident
measurements would also allow EMRIs to serve as standard sirens to
independently test cosmological
models~\cite{1986Natur.323..310S,*2005ApJ...629...15H}. LISA EMRIs are
low-redshift events, for which weak lensing errors, dominant
in comparable mass, SMBH standard sirens at higher-$z$,
are subdominant~\cite{2006ApJ...637...27K}.
Disk effects will not compromise the ability to constrain cosmological parameters,
as they  enter the GW observable with a different frequency signature, and are thus
weakly correlated. Migration effects may also deplete the unresolved
low-frequency EMRIs that contribute to the GW confusion noise
background~\cite{2004PhRvD..70l2002B}.

{\emph{Acknowledgements.}}---BK and NY acknowledge support from the
NASA through Einstein Fellowship PF9-00063 and PF0-110080 issued by
the Chandra X-ray Observatory, operated by the SAO,
on behalf of NASA under NAS8-03060. AL acknowledges support from
NSF grant AST-0907890 and NASA grants NNX08AL43G
and NNA09DB30A, and ZH from the Pol\'anyi Program of the Hungarian
National Office for Research and Technology (NKTH) and from NASA grant
NNX08AH35G.

\bibliography{review}

\begin{thebibliography}{43}%
\makeatletter
\providecommand \@ifxundefined [1]{%
 \@ifx{#1\undefined}
}%
\providecommand \@ifnum [1]{%
 \ifnum #1\expandafter \@firstoftwo
 \else \expandafter \@secondoftwo
 \fi
}%
\providecommand \@ifx [1]{%
 \ifx #1\expandafter \@firstoftwo
 \else \expandafter \@secondoftwo
 \fi
}%
\providecommand \natexlab [1]{#1}%
\providecommand \enquote  [1]{``#1''}%
\providecommand \bibnamefont  [1]{#1}%
\providecommand \bibfnamefont [1]{#1}%
\providecommand \citenamefont [1]{#1}%
\providecommand \href@noop [0]{\@secondoftwo}%
\providecommand \href [0]{\begingroup \@sanitize@url \@href}%
\providecommand \@href[1]{\@@startlink{#1}\@@href}%
\providecommand \@@href[1]{\endgroup#1\@@endlink}%
\providecommand \@sanitize@url [0]{\catcode `\\12\catcode `\$12\catcode
  `\&12\catcode `\#12\catcode `\^12\catcode `\_12\catcode `\%12\relax}%
\providecommand \@@startlink[1]{}%
\providecommand \@@endlink[0]{}%
\providecommand \url  [0]{\begingroup\@sanitize@url \@url }%
\providecommand \@url [1]{\endgroup\@href {#1}{\urlprefix }}%
\providecommand \urlprefix  [0]{URL }%
\providecommand \Eprint [0]{\href }%
\providecommand \doibase [0]{http://dx.doi.org/}%
\providecommand \selectlanguage [0]{\@gobble}%
\providecommand \bibinfo  [0]{\@secondoftwo}%
\providecommand \bibfield  [0]{\@secondoftwo}%
\providecommand \translation [1]{[#1]}%
\providecommand \BibitemOpen [0]{}%
\providecommand \bibitemStop [0]{}%
\providecommand \bibitemNoStop [0]{.\EOS\space}%
\providecommand \EOS [0]{\spacefactor3000\relax}%
\providecommand \BibitemShut  [1]{\csname bibitem#1\endcsname}%
\let\auto@bib@innerbib\@empty
\bibitem [{\citenamefont {LISA}()}]{lisa}%
  \BibitemOpen
  \bibfield  {author} {\bibinfo {author} {\bibnamefont {LISA}},\ }\href@noop {}
  {}\bibinfo {note} {{\tt www.esa.int/science/lisa}}\BibitemShut {NoStop}%
\bibitem [{\citenamefont {{Goodman}}\ and\ \citenamefont
  {{Tan}}(2004)}]{2004ApJ...608..108G}%
  \BibitemOpen
  \bibfield  {author} {\bibinfo {author} {\bibfnamefont {J.}~\bibnamefont
  {{Goodman}}}\ and\ \bibinfo {author} {\bibfnamefont {J.~C.}\ \bibnamefont
  {{Tan}}},\ }\href {\doibase 10.1086/386360} {\bibfield  {journal} {\bibinfo
  {journal} {\apj}\ }\textbf {\bibinfo {volume} {608}},\ \bibinfo {pages} {108}
  (\bibinfo {year} {2004})}\BibitemShut {NoStop}%
\bibitem [{\citenamefont {Levin}(2007)}]{Levin:2006uc}%
  \BibitemOpen
  \bibfield  {author} {\bibinfo {author} {\bibfnamefont {Y.}~\bibnamefont
  {Levin}},\ }\href {\doibase 10.1111/j.1365-2966.2006.11155.x} {\bibfield
  {journal} {\bibinfo  {journal} {Mon. Not. Roy. Astron. Soc.}\ }\textbf
  {\bibinfo {volume} {374}},\ \bibinfo {pages} {515} (\bibinfo {year}
  {2007})}\BibitemShut {NoStop}%
\bibitem [{\citenamefont {{{\v S}ubr}}\ and\ \citenamefont
  {{Karas}}(1999)}]{1999A&A...352..452S}%
  \BibitemOpen
  \bibfield  {author} {\bibinfo {author} {\bibfnamefont {L.}~\bibnamefont {{{\v
  S}ubr}}}\ and\ \bibinfo {author} {\bibfnamefont {V.}~\bibnamefont
  {{Karas}}},\ }\href@noop {} {\bibfield  {journal} {\bibinfo  {journal}
  {\aap}\ }\textbf {\bibinfo {volume} {352}},\ \bibinfo {pages} {452} (\bibinfo
  {year} {1999})}\BibitemShut {NoStop}%
\bibitem [{\citenamefont {{Miralda-Escud{\'e}}}\ and\ \citenamefont
  {{Kollmeier}}(2005)}]{2005ApJ...619...30M}%
  \BibitemOpen
  \bibfield  {author} {\bibinfo {author} {\bibfnamefont {J.}~\bibnamefont
  {{Miralda-Escud{\'e}}}}\ and\ \bibinfo {author} {\bibfnamefont {J.~A.}\
  \bibnamefont {{Kollmeier}}},\ }\href {\doibase 10.1086/426467} {\bibfield
  {journal} {\bibinfo  {journal} {\apj}\ }\textbf {\bibinfo {volume} {619}},\
  \bibinfo {pages} {30} (\bibinfo {year} {2005})}\BibitemShut {NoStop}%
\bibitem [{\citenamefont {{Freitag}}\ \emph {et~al.}(2006)\citenamefont
  {{Freitag}}, \citenamefont {{Amaro-Seoane}},\ and\ \citenamefont
  {{Kalogera}}}]{2006ApJ...649...91F}%
  \BibitemOpen
  \bibfield  {author} {\bibinfo {author} {\bibfnamefont {M.}~\bibnamefont
  {{Freitag}}}, \bibinfo {author} {\bibfnamefont {P.}~\bibnamefont
  {{Amaro-Seoane}}}, \ and\ \bibinfo {author} {\bibfnamefont {V.}~\bibnamefont
  {{Kalogera}}},\ }\href {\doibase 10.1086/506193} {\bibfield  {journal}
  {\bibinfo  {journal} {\apj}\ }\textbf {\bibinfo {volume} {649}},\ \bibinfo
  {pages} {91} (\bibinfo {year} {2006})}\BibitemShut {NoStop}%
\bibitem [{\citenamefont {{Amaro-Seoane}}\ \emph {et~al.}(2007)\citenamefont
  {{Amaro-Seoane}}, \citenamefont {{Gair}}, \citenamefont {{Freitag}},
  \citenamefont {{Miller}}, \citenamefont {{Mandel}}, \citenamefont
  {{Cutler}},\ and\ \citenamefont {{Babak}}}]{2007CQGra..24..113A}%
  \BibitemOpen
  \bibfield  {author} {\bibinfo {author} {\bibfnamefont {P.}~\bibnamefont
  {{Amaro-Seoane}}}, \bibinfo {author} {\bibfnamefont {J.~R.}\ \bibnamefont
  {{Gair}}}, \bibinfo {author} {\bibfnamefont {M.}~\bibnamefont {{Freitag}}},
  \bibinfo {author} {\bibfnamefont {M.~C.}\ \bibnamefont {{Miller}}}, \bibinfo
  {author} {\bibfnamefont {I.}~\bibnamefont {{Mandel}}}, \bibinfo {author}
  {\bibfnamefont {C.~J.}\ \bibnamefont {{Cutler}}}, \ and\ \bibinfo {author}
  {\bibfnamefont {S.}~\bibnamefont {{Babak}}},\ }\href {\doibase
  10.1088/0264-9381/24/17/R01} {\bibfield  {journal} {\bibinfo  {journal}
  {Classical and Quantum Gravity}\ }\textbf {\bibinfo {volume} {24}},\ \bibinfo
  {pages} {113} (\bibinfo {year} {2007})}\BibitemShut {NoStop}%
\bibitem [{\citenamefont {Gair}(2009)}]{Gair:2008bx}%
  \BibitemOpen
  \bibfield  {author} {\bibinfo {author} {\bibfnamefont {J.~R.}\ \bibnamefont
  {Gair}},\ }\href {\doibase 10.1088/0264-9381/26/9/094034} {\bibfield
  {journal} {\bibinfo  {journal} {Class. Quant. Grav.}\ }\textbf {\bibinfo
  {volume} {26}},\ \bibinfo {pages} {094034} (\bibinfo {year}
  {2009})}\BibitemShut {NoStop}%
\bibitem [{\citenamefont {{Armitage}}(2007)}]{2007astro.ph..1485A}%
  \BibitemOpen
  \bibfield  {author} {\bibinfo {author} {\bibfnamefont {P.~J.}\ \bibnamefont
  {{Armitage}}},\ }\href@noop {} {\  (\bibinfo {year} {2007})}\BibitemShut
  {NoStop}%
\bibitem [{\citenamefont {{Ward}}(1997)}]{1997Icar..126..261W}%
  \BibitemOpen
  \bibfield  {author} {\bibinfo {author} {\bibfnamefont {W.~R.}\ \bibnamefont
  {{Ward}}},\ }\href {\doibase 10.1006/icar.1996.5647} {\bibfield  {journal}
  {\bibinfo  {journal} {\icarus}\ }\textbf {\bibinfo {volume} {126}},\ \bibinfo
  {pages} {261} (\bibinfo {year} {1997})}\BibitemShut {NoStop}%
\bibitem [{\citenamefont {{Karas}}\ and\ \citenamefont {{{\v
  S}ubr}}(2001)}]{2001A&A...376..686K}%
  \BibitemOpen
  \bibfield  {author} {\bibinfo {author} {\bibfnamefont {V.}~\bibnamefont
  {{Karas}}}\ and\ \bibinfo {author} {\bibfnamefont {L.}~\bibnamefont {{{\v
  S}ubr}}},\ }\href {\doibase 10.1051/0004-6361:20011009} {\bibfield  {journal}
  {\bibinfo  {journal} {\aap}\ }\textbf {\bibinfo {volume} {376}},\ \bibinfo
  {pages} {686} (\bibinfo {year} {2001})}\BibitemShut {NoStop}%
\bibitem [{\citenamefont {Kocsis}\ \emph {et~al.}()\citenamefont {Kocsis},
  \citenamefont {Yunes},\ and\ \citenamefont {Loeb}}]{long-paper}%
  \BibitemOpen
  \bibfield  {author} {\bibinfo {author} {\bibfnamefont {B.}~\bibnamefont
  {Kocsis}}, \bibinfo {author} {\bibfnamefont {N.}~\bibnamefont {Yunes}}, \
  and\ \bibinfo {author} {\bibfnamefont {A.}~\bibnamefont {Loeb}},\ }\href@noop
  {} {}\bibinfo {note} {In preparation (2011)}\BibitemShut {NoStop}%
\bibitem [{\citenamefont {{Kollmeier~et~al.}}(2006)}]{2006ApJ...648..128K}%
  \BibitemOpen
  \bibfield  {author} {\bibinfo {author} {\bibfnamefont {J.~A.}\ \bibnamefont
  {{Kollmeier~et~al.}}},\ }\href {\doibase 10.1086/505646} {\bibfield
  {journal} {\bibinfo  {journal} {\apj}\ }\textbf {\bibinfo {volume} {648}},\
  \bibinfo {pages} {128} (\bibinfo {year} {2006})}\BibitemShut {NoStop}%
\bibitem [{\citenamefont {{Pessah}}\ \emph {et~al.}(2007)\citenamefont
  {{Pessah}}, \citenamefont {{Chan}},\ and\ \citenamefont
  {{Psaltis}}}]{2007ApJ...668L..51P}%
  \BibitemOpen
  \bibfield  {author} {\bibinfo {author} {\bibfnamefont {M.~E.}\ \bibnamefont
  {{Pessah}}}, \bibinfo {author} {\bibfnamefont {C.}~\bibnamefont {{Chan}}}, \
  and\ \bibinfo {author} {\bibfnamefont {D.}~\bibnamefont {{Psaltis}}},\ }\href
  {\doibase 10.1086/522585} {\bibfield  {journal} {\bibinfo  {journal} {\apjl}\
  }\textbf {\bibinfo {volume} {668}},\ \bibinfo {pages} {L51} (\bibinfo {year}
  {2007})}\BibitemShut {NoStop}%
\bibitem [{\citenamefont {{Dubus}}\ \emph {et~al.}(2001)\citenamefont
  {{Dubus}}, \citenamefont {{Hameury}},\ and\ \citenamefont
  {{Lasota}}}]{2001A&A...373..251D}%
  \BibitemOpen
  \bibfield  {author} {\bibinfo {author} {\bibfnamefont {G.}~\bibnamefont
  {{Dubus}}}, \bibinfo {author} {\bibfnamefont {J.}~\bibnamefont {{Hameury}}},
  \ and\ \bibinfo {author} {\bibfnamefont {J.}~\bibnamefont {{Lasota}}},\
  }\href {\doibase 10.1051/0004-6361:20010632} {\bibfield  {journal} {\bibinfo
  {journal} {\aap}\ }\textbf {\bibinfo {volume} {373}},\ \bibinfo {pages} {251}
  (\bibinfo {year} {2001})}\BibitemShut {NoStop}%
\bibitem [{\citenamefont {{King}}\ \emph {et~al.}(2007)\citenamefont {{King}},
  \citenamefont {{Pringle}},\ and\ \citenamefont
  {{Livio}}}]{2007MNRAS.376.1740K}%
  \BibitemOpen
  \bibfield  {author} {\bibinfo {author} {\bibfnamefont {A.~R.}\ \bibnamefont
  {{King}}}, \bibinfo {author} {\bibfnamefont {J.~E.}\ \bibnamefont
  {{Pringle}}}, \ and\ \bibinfo {author} {\bibfnamefont {M.}~\bibnamefont
  {{Livio}}},\ }\href {\doibase 10.1111/j.1365-2966.2007.11556.x} {\bibfield
  {journal} {\bibinfo  {journal} {\mnras}\ }\textbf {\bibinfo {volume} {376}},\
  \bibinfo {pages} {1740} (\bibinfo {year} {2007})}\BibitemShut {NoStop}%
\bibitem [{\citenamefont {{Shakura}}\ and\ \citenamefont
  {{Sunyaev}}(1973)}]{1973A&A....24..337S}%
  \BibitemOpen
  \bibfield  {author} {\bibinfo {author} {\bibfnamefont {N.~I.}\ \bibnamefont
  {{Shakura}}}\ and\ \bibinfo {author} {\bibfnamefont {R.~A.}\ \bibnamefont
  {{Sunyaev}}},\ }\href@noop {} {\bibfield  {journal} {\bibinfo  {journal}
  {Astron. Astroph.}\ }\textbf {\bibinfo {volume} {24}},\ \bibinfo {pages}
  {337} (\bibinfo {year} {1973})}\BibitemShut {NoStop}%
\bibitem [{\citenamefont {{Sakimoto}}\ and\ \citenamefont
  {{Coroniti}}(1981)}]{1981ApJ...247...19S}%
  \BibitemOpen
  \bibfield  {author} {\bibinfo {author} {\bibfnamefont {P.~J.}\ \bibnamefont
  {{Sakimoto}}}\ and\ \bibinfo {author} {\bibfnamefont {F.~V.}\ \bibnamefont
  {{Coroniti}}},\ }\href {\doibase 10.1086/159005} {\bibfield  {journal}
  {\bibinfo  {journal} {\apj}\ }\textbf {\bibinfo {volume} {247}},\ \bibinfo
  {pages} {19} (\bibinfo {year} {1981})}\BibitemShut {NoStop}%
\bibitem [{\citenamefont {{Goldreich}}\ and\ \citenamefont
  {{Tremaine}}(1980)}]{1980ApJ...241..425G}%
  \BibitemOpen
  \bibfield  {author} {\bibinfo {author} {\bibfnamefont {P.}~\bibnamefont
  {{Goldreich}}}\ and\ \bibinfo {author} {\bibfnamefont {S.}~\bibnamefont
  {{Tremaine}}},\ }\href {\doibase 10.1086/158356} {\bibfield  {journal}
  {\bibinfo  {journal} {\apj}\ }\textbf {\bibinfo {volume} {241}},\ \bibinfo
  {pages} {425} (\bibinfo {year} {1980})}\BibitemShut {NoStop}%
\bibitem [{\citenamefont {{Tanaka}}\ \emph {et~al.}(2002)\citenamefont
  {{Tanaka}}, \citenamefont {{Takeuchi}},\ and\ \citenamefont
  {{Ward}}}]{2002ApJ...565.1257T}%
  \BibitemOpen
  \bibfield  {author} {\bibinfo {author} {\bibfnamefont {H.}~\bibnamefont
  {{Tanaka}}}, \bibinfo {author} {\bibfnamefont {T.}~\bibnamefont
  {{Takeuchi}}}, \ and\ \bibinfo {author} {\bibfnamefont {W.~R.}\ \bibnamefont
  {{Ward}}},\ }\href {\doibase 10.1086/324713} {\bibfield  {journal} {\bibinfo
  {journal} {\apj}\ }\textbf {\bibinfo {volume} {565}},\ \bibinfo {pages}
  {1257} (\bibinfo {year} {2002})}\BibitemShut {NoStop}%
\bibitem [{\citenamefont {{Paardekooper}}\ and\ \citenamefont
  {{Mellema}}(2006)}]{2006A&A...459L..17P}%
  \BibitemOpen
  \bibfield  {author} {\bibinfo {author} {\bibfnamefont {S.}~\bibnamefont
  {{Paardekooper}}}\ and\ \bibinfo {author} {\bibfnamefont {G.}~\bibnamefont
  {{Mellema}}},\ }\href {\doibase 10.1051/0004-6361:20066304} {\bibfield
  {journal} {\bibinfo  {journal} {\aap}\ }\textbf {\bibinfo {volume} {459}},\
  \bibinfo {pages} {L17} (\bibinfo {year} {2006})}\BibitemShut {NoStop}%
\bibitem [{\citenamefont {{Nelson}}\ and\ \citenamefont
  {{Papaloizou}}(2004)}]{2004MNRAS.350..849N}%
  \BibitemOpen
  \bibfield  {author} {\bibinfo {author} {\bibfnamefont {R.~P.}\ \bibnamefont
  {{Nelson}}}\ and\ \bibinfo {author} {\bibfnamefont {J.~C.~B.}\ \bibnamefont
  {{Papaloizou}}},\ }\href {\doibase 10.1111/j.1365-2966.2004.07406.x}
  {\bibfield  {journal} {\bibinfo  {journal} {\mnras}\ }\textbf {\bibinfo
  {volume} {350}},\ \bibinfo {pages} {849} (\bibinfo {year}
  {2004})}\BibitemShut {NoStop}%
\bibitem [{\citenamefont {{Laughlin}}\ \emph {et~al.}(2004)\citenamefont
  {{Laughlin}}, \citenamefont {{Steinacker}},\ and\ \citenamefont
  {{Adams}}}]{2004ApJ...608..489L}%
  \BibitemOpen
  \bibfield  {author} {\bibinfo {author} {\bibfnamefont {G.}~\bibnamefont
  {{Laughlin}}}, \bibinfo {author} {\bibfnamefont {A.}~\bibnamefont
  {{Steinacker}}}, \ and\ \bibinfo {author} {\bibfnamefont {F.~C.}\
  \bibnamefont {{Adams}}},\ }\href {\doibase 10.1086/386316} {\bibfield
  {journal} {\bibinfo  {journal} {\apj}\ }\textbf {\bibinfo {volume} {608}},\
  \bibinfo {pages} {489} (\bibinfo {year} {2004})}\BibitemShut {NoStop}%
\bibitem [{\citenamefont {{Syer}}\ and\ \citenamefont
  {{Clarke}}(1995)}]{1995MNRAS.277..758S}%
  \BibitemOpen
  \bibfield  {author} {\bibinfo {author} {\bibfnamefont {D.}~\bibnamefont
  {{Syer}}}\ and\ \bibinfo {author} {\bibfnamefont {C.~J.}\ \bibnamefont
  {{Clarke}}},\ }\href@noop {} {\bibfield  {journal} {\bibinfo  {journal}
  {\mnras}\ }\textbf {\bibinfo {volume} {277}},\ \bibinfo {pages} {758}
  (\bibinfo {year} {1995})}\BibitemShut {NoStop}%
\bibitem [{\citenamefont {{Ivanov}}\ \emph {et~al.}(1999)\citenamefont
  {{Ivanov}}, \citenamefont {{Papaloizou}},\ and\ \citenamefont
  {{Polnarev}}}]{1999MNRAS.307...79I}%
  \BibitemOpen
  \bibfield  {author} {\bibinfo {author} {\bibfnamefont {P.~B.}\ \bibnamefont
  {{Ivanov}}}, \bibinfo {author} {\bibfnamefont {J.~C.~B.}\ \bibnamefont
  {{Papaloizou}}}, \ and\ \bibinfo {author} {\bibfnamefont {A.~G.}\
  \bibnamefont {{Polnarev}}},\ }\href {\doibase
  10.1046/j.1365-8711.1999.02623.x} {\bibfield  {journal} {\bibinfo  {journal}
  {\mnras}\ }\textbf {\bibinfo {volume} {307}},\ \bibinfo {pages} {79}
  (\bibinfo {year} {1999})}\BibitemShut {NoStop}%
\bibitem [{\citenamefont {{Milosavljevi{\'c}}}\ and\ \citenamefont
  {{Phinney}}(2005)}]{2005ApJ...622L..93M}%
  \BibitemOpen
  \bibfield  {author} {\bibinfo {author} {\bibfnamefont {M.}~\bibnamefont
  {{Milosavljevi{\'c}}}}\ and\ \bibinfo {author} {\bibfnamefont {E.~S.}\
  \bibnamefont {{Phinney}}},\ }\href {\doibase 10.1086/429618} {\bibfield
  {journal} {\bibinfo  {journal} {\apjl}\ }\textbf {\bibinfo {volume} {622}},\
  \bibinfo {pages} {L93} (\bibinfo {year} {2005})}\BibitemShut {NoStop}%
\bibitem [{\citenamefont {{Haiman}}\ \emph {et~al.}(2009)\citenamefont
  {{Haiman}}, \citenamefont {{Kocsis}},\ and\ \citenamefont
  {{Menou}}}]{2009ApJ...700.1952H}%
  \BibitemOpen
  \bibfield  {author} {\bibinfo {author} {\bibfnamefont {Z.}~\bibnamefont
  {{Haiman}}}, \bibinfo {author} {\bibfnamefont {B.}~\bibnamefont {{Kocsis}}},
  \ and\ \bibinfo {author} {\bibfnamefont {K.}~\bibnamefont {{Menou}}},\ }\href
  {\doibase 10.1088/0004-637X/700/2/1952} {\bibfield  {journal} {\bibinfo
  {journal} {\apj}\ }\textbf {\bibinfo {volume} {700}},\ \bibinfo {pages}
  {1952} (\bibinfo {year} {2009})}\BibitemShut {NoStop}%
\bibitem [{\citenamefont {{Artymowicz}}\ and\ \citenamefont
  {{Lubow}}(1996)}]{1996ApJ...467L..77A}%
  \BibitemOpen
  \bibfield  {author} {\bibinfo {author} {\bibfnamefont {P.}~\bibnamefont
  {{Artymowicz}}}\ and\ \bibinfo {author} {\bibfnamefont {S.~H.}\ \bibnamefont
  {{Lubow}}},\ }\href {\doibase 10.1086/310200} {\bibfield  {journal} {\bibinfo
   {journal} {\apjl}\ }\textbf {\bibinfo {volume} {467}},\ \bibinfo {pages}
  {L77+} (\bibinfo {year} {1996})}\BibitemShut {NoStop}%
\bibitem [{\citenamefont {{MacFadyen}}\ and\ \citenamefont
  {{Milosavljevi{\'c}}}(2008)}]{2008ApJ...672...83M}%
  \BibitemOpen
  \bibfield  {author} {\bibinfo {author} {\bibfnamefont {A.~I.}\ \bibnamefont
  {{MacFadyen}}}\ and\ \bibinfo {author} {\bibfnamefont {M.}~\bibnamefont
  {{Milosavljevi{\'c}}}},\ }\href {\doibase 10.1086/523869} {\bibfield
  {journal} {\bibinfo  {journal} {\apj}\ }\textbf {\bibinfo {volume} {672}},\
  \bibinfo {pages} {83} (\bibinfo {year} {2008})}\BibitemShut {NoStop}%
\bibitem [{\citenamefont {{Cuadra}}\ \emph {et~al.}(2009)\citenamefont
  {{Cuadra}}, \citenamefont {{Armitage}}, \citenamefont {{Alexander}},\ and\
  \citenamefont {{Begelman}}}]{2009MNRAS.393.1423C}%
  \BibitemOpen
  \bibfield  {author} {\bibinfo {author} {\bibfnamefont {J.}~\bibnamefont
  {{Cuadra}}}, \bibinfo {author} {\bibfnamefont {P.~J.}\ \bibnamefont
  {{Armitage}}}, \bibinfo {author} {\bibfnamefont {R.~D.}\ \bibnamefont
  {{Alexander}}}, \ and\ \bibinfo {author} {\bibfnamefont {M.~C.}\ \bibnamefont
  {{Begelman}}},\ }\href {\doibase 10.1111/j.1365-2966.2008.14147.x} {\bibfield
   {journal} {\bibinfo  {journal} {\mnras}\ }\textbf {\bibinfo {volume}
  {393}},\ \bibinfo {pages} {1423} (\bibinfo {year} {2009})}\BibitemShut
  {NoStop}%
\bibitem [{\citenamefont {{Artymowicz}}\ and\ \citenamefont
  {{Lubow}}(1994)}]{1994ApJ...421..651A}%
  \BibitemOpen
  \bibfield  {author} {\bibinfo {author} {\bibfnamefont {P.}~\bibnamefont
  {{Artymowicz}}}\ and\ \bibinfo {author} {\bibfnamefont {S.~H.}\ \bibnamefont
  {{Lubow}}},\ }\href {\doibase 10.1086/173679} {\bibfield  {journal} {\bibinfo
   {journal} {\apj}\ }\textbf {\bibinfo {volume} {421}},\ \bibinfo {pages}
  {651} (\bibinfo {year} {1994})}\BibitemShut {NoStop}%
\bibitem [{\citenamefont {{Barack}}\ and\ \citenamefont
  {{Cutler}}(2004)}]{2004PhRvD..70l2002B}%
  \BibitemOpen
  \bibfield  {author} {\bibinfo {author} {\bibfnamefont {L.}~\bibnamefont
  {{Barack}}}\ and\ \bibinfo {author} {\bibfnamefont {C.}~\bibnamefont
  {{Cutler}}},\ }\href {\doibase 10.1103/PhysRevD.70.122002} {\bibfield
  {journal} {\bibinfo  {journal} {\prd}\ }\textbf {\bibinfo {volume} {70}},\
  \bibinfo {pages} {122002} (\bibinfo {year} {2004})}\BibitemShut {NoStop}%
\bibitem [{\citenamefont {{Yunes}}\ \emph {et~al.}(2010)\citenamefont
  {{Yunes}}, \citenamefont {{Buonanno}}, \citenamefont {{Hughes}},
  \citenamefont {{Miller}},\ and\ \citenamefont {{Pan}}}]{Yunes:2009ef}%
  \BibitemOpen
  \bibfield  {author} {\bibinfo {author} {\bibfnamefont {N.}~\bibnamefont
  {{Yunes}}}, \bibinfo {author} {\bibfnamefont {A.}~\bibnamefont {{Buonanno}}},
  \bibinfo {author} {\bibfnamefont {S.~A.}\ \bibnamefont {{Hughes}}}, \bibinfo
  {author} {\bibfnamefont {M.~C.}\ \bibnamefont {{Miller}}}, \ and\ \bibinfo
  {author} {\bibfnamefont {Y.}~\bibnamefont {{Pan}}},\ }\href {\doibase
  10.1103/PhysRevLett.104.091102} {\bibfield  {journal} {\bibinfo  {journal}
  {Physical Review Letters}\ }\textbf {\bibinfo {volume} {104}},\ \bibinfo
  {pages} {091102} (\bibinfo {year} {2010})}\BibitemShut {NoStop}%
\bibitem [{\citenamefont {{Yunes}}(2009)}]{2009GWN.....2....3Y}%
  \BibitemOpen
  \bibfield  {author} {\bibinfo {author} {\bibfnamefont {N.}~\bibnamefont
  {{Yunes}}},\ }\href@noop {} {\bibfield  {journal} {\bibinfo  {journal} {GW
  Notes, Vol.~2, p.~3-47}\ }\textbf {\bibinfo {volume} {2}},\ \bibinfo {pages}
  {3} (\bibinfo {year} {2009})}\BibitemShut {NoStop}%
\bibitem [{\citenamefont {Yunes}\ \emph
  {et~al.}(2010{\natexlab{a}})\citenamefont {Yunes}, \citenamefont {Buonanno},
  \citenamefont {Hughes}, \citenamefont {Pan}, \citenamefont {Barausse} \emph
  {et~al.}}]{Yunes:2010zj}%
  \BibitemOpen
  \bibfield  {author} {\bibinfo {author} {\bibfnamefont {N.}~\bibnamefont
  {Yunes}}, \bibinfo {author} {\bibfnamefont {A.}~\bibnamefont {Buonanno}},
  \bibinfo {author} {\bibfnamefont {S.~A.}\ \bibnamefont {Hughes}}, \bibinfo
  {author} {\bibfnamefont {Y.}~\bibnamefont {Pan}}, \bibinfo {author}
  {\bibfnamefont {E.}~\bibnamefont {Barausse}},  \emph {et~al.},\ }\href@noop
  {} {\  (\bibinfo {year} {2010}{\natexlab{a}})}\BibitemShut {NoStop}%
\bibitem [{\citenamefont {{Poisson}}\ and\ \citenamefont
  {{Will}}(1995)}]{1995PhRvD..52..848P}%
  \BibitemOpen
  \bibfield  {author} {\bibinfo {author} {\bibfnamefont {E.}~\bibnamefont
  {{Poisson}}}\ and\ \bibinfo {author} {\bibfnamefont {C.~M.}\ \bibnamefont
  {{Will}}},\ }\href {\doibase 10.1103/PhysRevD.52.848} {\bibfield  {journal}
  {\bibinfo  {journal} {\prd}\ }\textbf {\bibinfo {volume} {52}},\ \bibinfo
  {pages} {848} (\bibinfo {year} {1995})}\BibitemShut {NoStop}%
\bibitem [{\citenamefont {Yunes}\ \emph {et~al.}(2009)\citenamefont {Yunes},
  \citenamefont {Arun}, \citenamefont {Berti},\ and\ \citenamefont
  {Will}}]{Yunes:2009yz}%
  \BibitemOpen
  \bibfield  {author} {\bibinfo {author} {\bibfnamefont {N.}~\bibnamefont
  {Yunes}}, \bibinfo {author} {\bibfnamefont {K.}~\bibnamefont {Arun}},
  \bibinfo {author} {\bibfnamefont {E.}~\bibnamefont {Berti}}, \ and\ \bibinfo
  {author} {\bibfnamefont {C.~M.}\ \bibnamefont {Will}},\ }\href {\doibase
  10.1103/PhysRevD.80.084001} {\bibfield  {journal} {\bibinfo  {journal}
  {Phys.Rev.}\ }\textbf {\bibinfo {volume} {D80}},\ \bibinfo {pages} {084001}
  (\bibinfo {year} {2009})}\BibitemShut {NoStop}%
\bibitem [{\citenamefont {Yunes}\ and\ \citenamefont
  {Pretorius}(2009)}]{Yunes:2009ke}%
  \BibitemOpen
  \bibfield  {author} {\bibinfo {author} {\bibfnamefont {N.}~\bibnamefont
  {Yunes}}\ and\ \bibinfo {author} {\bibfnamefont {F.}~\bibnamefont
  {Pretorius}},\ }\href {\doibase 10.1103/PhysRevD.80.122003} {\bibfield
  {journal} {\bibinfo  {journal} {Phys.Rev.}\ }\textbf {\bibinfo {volume}
  {D80}},\ \bibinfo {pages} {122003} (\bibinfo {year} {2009})}\BibitemShut
  {NoStop}%
\bibitem [{\citenamefont {Yunes}\ \emph
  {et~al.}(2010{\natexlab{b}})\citenamefont {Yunes}, \citenamefont
  {Pretorius},\ and\ \citenamefont {Spergel}}]{Yunes:2009bv}%
  \BibitemOpen
  \bibfield  {author} {\bibinfo {author} {\bibfnamefont {N.}~\bibnamefont
  {Yunes}}, \bibinfo {author} {\bibfnamefont {F.}~\bibnamefont {Pretorius}}, \
  and\ \bibinfo {author} {\bibfnamefont {D.}~\bibnamefont {Spergel}},\ }\href
  {\doibase 10.1103/PhysRevD.81.064018} {\bibfield  {journal} {\bibinfo
  {journal} {Phys.Rev.}\ }\textbf {\bibinfo {volume} {D81}},\ \bibinfo {pages}
  {064018} (\bibinfo {year} {2010}{\natexlab{b}})}\BibitemShut {NoStop}%
\bibitem [{\citenamefont {{Kocsis}}\ \emph {et~al.}(2006)\citenamefont
  {{Kocsis}}, \citenamefont {{Frei}}, \citenamefont {{Haiman}},\ and\
  \citenamefont {{Menou}}}]{2006ApJ...637...27K}%
  \BibitemOpen
  \bibfield  {author} {\bibinfo {author} {\bibfnamefont {B.}~\bibnamefont
  {{Kocsis}}}, \bibinfo {author} {\bibfnamefont {Z.}~\bibnamefont {{Frei}}},
  \bibinfo {author} {\bibfnamefont {Z.}~\bibnamefont {{Haiman}}}, \ and\
  \bibinfo {author} {\bibfnamefont {K.}~\bibnamefont {{Menou}}},\ }\href
  {\doibase 10.1086/498236} {\bibfield  {journal} {\bibinfo  {journal} {\apj}\
  }\textbf {\bibinfo {volume} {637}},\ \bibinfo {pages} {27} (\bibinfo {year}
  {2006})}\BibitemShut {NoStop}%
\bibitem [{\citenamefont {{Kocsis}}\ \emph {et~al.}(2008)\citenamefont
  {{Kocsis}}, \citenamefont {{Haiman}},\ and\ \citenamefont
  {{Menou}}}]{2008ApJ...684..870K}%
  \BibitemOpen
  \bibfield  {author} {\bibinfo {author} {\bibfnamefont {B.}~\bibnamefont
  {{Kocsis}}}, \bibinfo {author} {\bibfnamefont {Z.}~\bibnamefont {{Haiman}}},
  \ and\ \bibinfo {author} {\bibfnamefont {K.}~\bibnamefont {{Menou}}},\ }\href
  {\doibase 10.1086/590230} {\bibfield  {journal} {\bibinfo  {journal} {\apj}\
  }\textbf {\bibinfo {volume} {684}},\ \bibinfo {pages} {870} (\bibinfo {year}
  {2008})}\BibitemShut {NoStop}%
\bibitem [{\citenamefont {{Schutz}}(1986)}]{1986Natur.323..310S}%
  \BibitemOpen
  \bibfield  {author} {\bibinfo {author} {\bibfnamefont {B.~F.}\ \bibnamefont
  {{Schutz}}},\ }\href {\doibase 10.1038/323310a0} {\bibfield  {journal}
  {\bibinfo  {journal} {\nat}\ }\textbf {\bibinfo {volume} {323}},\ \bibinfo
  {pages} {310} (\bibinfo {year} {1986})}\BibitemShut {NoStop}%
\bibitem [{\citenamefont {{Holz}}\ and\ \citenamefont
  {{Hughes}}(2005)}]{2005ApJ...629...15H}%
  \BibitemOpen
  \bibfield  {author} {\bibinfo {author} {\bibfnamefont {D.~E.}\ \bibnamefont
  {{Holz}}}\ and\ \bibinfo {author} {\bibfnamefont {S.~A.}\ \bibnamefont
  {{Hughes}}},\ }\href {\doibase 10.1086/431341} {\bibfield  {journal}
  {\bibinfo  {journal} {\apj}\ }\textbf {\bibinfo {volume} {629}},\ \bibinfo
  {pages} {15} (\bibinfo {year} {2005})}\BibitemShut {NoStop}%
\end{thebibliography}%
\end{document}